\documentclass[11pt,a4paper]{article}
\usepackage[hyperref]{acl2019}
\usepackage{times}
\usepackage{latexsym}

\usepackage{url}

\aclfinalcopy % Uncomment this line for the final submission

\usepackage[T1]{fontenc}

\usepackage{latexsym}

\usepackage{amsmath}
\usepackage{bbm}

\usepackage{amssymb}
\usepackage{algorithm}
\usepackage{algpseudocode}
\usepackage{multirow}
\usepackage{tabu}
\usepackage{booktabs}

\usepackage{graphicx} % more modern
\usepackage{subfigure}
\usepackage{booktabs}       % professional-quality tables
\usepackage{breakcites}

\usepackage{tikz}
\usepackage{pgfplots}
\pgfplotsset{compat=1.14}

\newcommand\ignore[1]{}

\usepackage{array}
\newcolumntype{H}{>{\setbox0=\hbox\bgroup}c<{\egroup}@{}}

\definecolor{attnred}{RGB}{242, 140, 142}
\newcommand\attn[2]{\fboxsep1pt \colorbox{attnred!#2}{\strut #1}}

\newcolumntype{C}[1]{>{\centering\let\newline\\\arraybackslash\hspace{0pt}}m{#1}}

\definecolor{g-blue}{HTML}{2E86C1}
\definecolor{g-red}{HTML}{B03A2E}
\definecolor{g-purple}{HTML}{AF7AC5}

\title{Multi-Stage Document Ranking with BERT}

\author{Rodrigo Nogueira,$^1$ Wei Yang,$^2$ Kyunghyun Cho,$^{3,4,5,6}$ \and Jimmy Lin$^2$\vspace{0.1cm}\\
$^1$ Tandon School of Engineering, New York University \\
$^2$ David R. Cheriton School of Computer Science, University of Waterloo \\
$^3$ Courant Institute of Mathematical Sciences, New York University \\
$^4$ Center for Data Science, New York University \\
$^5$ Facebook AI Research~~
$^6$ CIFAR Azrieli Global Scholar \\
}

\date{}

\begin{document}
\maketitle

\begin{abstract}
The advent of deep neural networks pre-trained via language modeling tasks has spurred a number of successful applications in natural language processing.
This work explores one such popular model, BERT, in the context of document ranking.
We propose two variants, called monoBERT and duoBERT, that formulate the ranking problem as pointwise and pairwise classification, respectively.
These two models are arranged in a multi-stage ranking architecture to form an end-to-end search system.
One major advantage of this design is the ability to trade off quality against latency by controlling the admission of candidates into each pipeline stage, and by doing so, we are able to find operating points that offer a good balance between these two competing metrics.
On two large-scale datasets, MS MARCO and TREC CAR, experiments show that our model produces results that are either at or comparable to the state of the art.
Ablation studies show the contributions of each component and characterize the latency/quality tradeoff space.
\end{abstract}

\section{Introduction}

Neural models pre-trained on language modeling tasks such as ELMo~\citep{peters2017semi}, Open\-AI GPT~\citep{radford2018improving}, and BERT~\citep{devlin2018bert} have achieved impressive results on NLP tasks ranging from natural language inference to question answering.
One such popular model, BERT, has recently been applied to search-related tasks, retrieval-based question answering~\cite{Yang_etal_NAACL2019demo}, as well as document ranking~\cite{Yang_etal_arXiv2019b,MacAvaney_etal_SIGIR2019,Yilmaz_etal_EMNLP2019}.

This paper builds on previous initial work~\citep{nogueira2019passage} to tackle the document ranking problem with a multi-stage ranking architecture.
We introduce two BERT variants, called monoBERT and duoBERT.
The monoBERT model treats document ranking as a binary classification problem over individual candidate documents, while the duoBERT model adopts a pairwise classification approach that considers pairs of candidate documents.
For end-to-end document ranking, we arrange these models as stages in a pipeline where each balances the size of the candidate set against the inherent complexity of the model. 
This design allows us to obtain the benefits of richer models while controlling the increased inference latencies that come with these richer models.

Our work makes the following contributions:
We start by describing monoBERT, a pointwise classification model of document relevance that was introduced in~\citet{nogueira2019passage}.
Second, we propose a novel extension of monoBERT, called duoBERT, that adopts a pairwise classification approach to document relevance.
Third, we integrate monoBERT and duoBERT in a multi-stage ranking architecture that allows us to reap the benefits of our richer duoBERT model with only a modest increase in inference latency.
The architecture adopts an innovation from the information retrieval (IR) community that to our knowledge has not been explored by NLP researchers.
Fourth, perhaps unsurprising, we show that pre-training on the corpus of the target task improves effectiveness over pre-training on out-of-domain corpora.

We evaluate our models on two large-scale document retrieval datasets that are conducive to deep learning experiments:\ the MS MARCO dataset and the Complex Answer Retrieval (CAR) Task at TREC.
On both datasets, our results are either at or comparable to the state of the art.
As we show through component-level ablation studies, both monoBERT and duoBERT contribute significantly to overall effectiveness.
Additionally, within the framework of multi-stage ranking, we characterize the latency vs.\ effectiveness tradeoff space of each model.

\section{Background and Related Work}

In this paper, we tackle the document ranking problem (also known as {\it ad hoc} retrieval), following the widely-accepted standard formulation:\
Given a user's information need expressed as a query $q$ and a (potentially large) corpus of documents, the system's task is to produce a ranking of $k$ documents that maximizes some metric, such as mean average precision (MAP) or mean reciprocal rank (MRR).
Throughout this paper, per standard parlance in IR, {\it document} is used generically to refer a unit of text being retrieved, when in actuality it may be a passage, a sentence, etc.

The basic idea behind multi-stage ranking is to break document ranking down into a series of pipeline stages.
Following an initial retrieval stage, which typically issues a ``bag of words'' query against an inverted index, each subsequent stage re-ranks the set of candidates passed along from the previous stage until the final output is returned to the user.
This basic approach has received much interest in academia~\cite{Matveeva_etal_SIGIR2006,Wang_etal_SIGIR2011,Asadi_Lin_SIGIR2013,ChenRuey-Cheng_etal_SIGIR2017a,Mackenzie_etal_WSDM2018} as well as industry.
Known production deployments include the Bing web search engine~\cite{Pedersen_SIGIR2010} as well as Alibaba's e-commerce search engine~\cite{LiuShichen_etal_SIGKDD2017}.

Multi-stage ranking architectures have evolved to strike a balance between model complexity and search latency by controlling the size of the candidate set at each stage.
Increasingly richer models can be made {\it practical} by considering successively smaller candidate sets.
For certain (easy) queries, stages of the pipeline can be skipped entirely, known as ``early exits''~\cite{Cambazoglu_etal_WSDM2010}.
Viewed in this manner, multi-stage ranking captures the same intuition as progressive refinement in classifier cascades~\cite{Viola_Jones_208}.
For example, an early stage might consider only term statistics of single terms, whereas later stages might consider bigrams, phrases, or even apply lightweight NLP techniques.
Given this setup, a number of researchers have proposed techniques based, for example, on boosting for composing these stages in an end-to-end manner~\cite{Wang_etal_SIGIR2011,XuZhixiang_etal_ICML2012}.
In our work, we make the connection between BERT-based models and multi-stage ranking, which allows us to trade off the quality of the results with inference latency.

The advent of deep learning has brought tremendous excitement into the information retrieval community.
Although machine-learned ranking models have been well studied since the mid-2000s under the banner of ``learning to rank'', the paradigm is heavily driven by manual feature engineering~\cite{LiuTY_FnTIR2009,LiHang_2011}; commercial web search engines are known to incorporate thousands of features (or more) in their models.
Continuous vector space representations coupled with neural models promise to obviate the need for handcrafted features and have attracted the attention of many researchers.
Well-known neural ranking models include DRMM~\citep{guo2016deep}, DUET~\citep{mitra2017learning}, KNRM~\citep{xiong2017end}, and Co-PACRR~\citep{hui2018co}; the literature is too vast for an exhaustive review here, and thus we refer readers to recent overviews~\cite{Onal_etal_IRJ2018,MitraBhaskar_Craswell_2019}.

Although often glossed over, most neural ranking models today (including all the models referenced above) are actually {\it re-ranking} models, in the sense that they operate over the output of a list of candidate documents, typically produced by a ``bag of words'' query.
Thus, document retrieval with neural models today already uses multi-stage ranking, albeit an impoverished form with only a single re-ranking stage.
This recognition provides a starting point of our work, from which we build BERT-based multi-stage ranking.

\begin{figure*}[ht]
\begin{center}
\centerline{\includegraphics[width=0.75\textwidth]{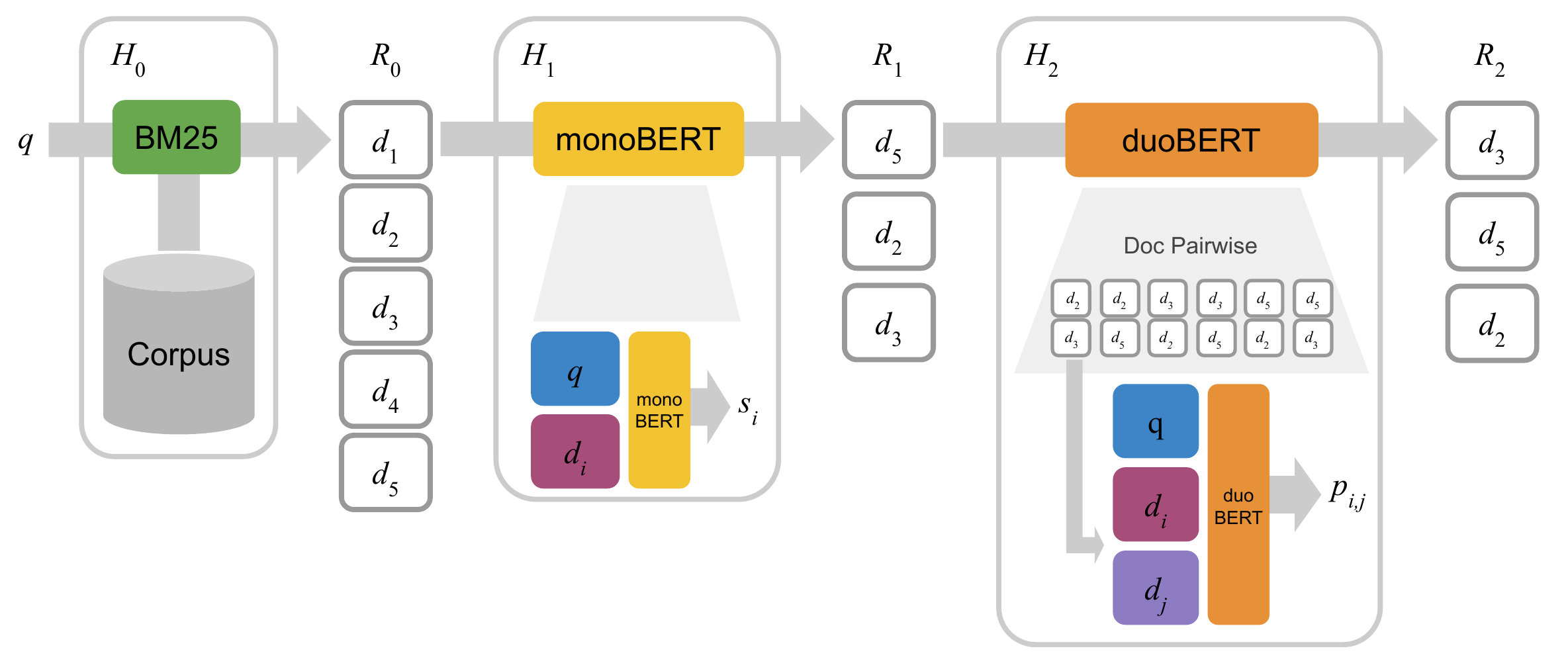}}
\caption{Illustration of our multi-stage ranking architecture. In the first stage $H_0$, given a query $q$, the top-$k_0$ ($k_0=5$ in the figure) candidate documents $R_0$ are retrieved using BM25. In the second stage $H_1$, monoBERT produces a relevance score $s_i$ for each pair of query $q$ and candidate $d_i \in R_0$. The top-$k_1$ ($k_1=3$ in the figure) candidates with respect to these relevance scores are passed to the last stage $H_2$, in which duoBERT computes a relevance score $p_{i,j}$ for each triple ($q$, $d_i$, $d_j$). The final list of candidates $R_2$ is formed by re-ranking the candidates according to these scores (see Section~\ref{sec:duoBERT_method} for a description of how these pairwise scores are aggregated).} 
\label{fig:overview}
\end{center}
\vspace{-5mm}
\end{figure*}

\section{Multi-Stage Ranking with BERT}

In our formulation, a multi-stage ranking architecture comprises a number of stages, denoted $H_0$ to $H_N$.
Except for $H_0$, which retrieves $k_0$ candidates from an inverted index, each stage $H_n$ receives a ranked list $R_{n-1}$ comprising $k_{n-1}$ candidates from the previous stage.
Each stage, in turn, provides a ranked list $R_n$ comprising $k_n$ candidates to the subsequent stage, with the obvious requirement that $k_n \le k_{n-1}$.
The ranked list generated by the final stage $H_N$ is designated for consumption by the (human) searcher.

For expository purposes, we consider stages to receive and produce {\it candidates} even though they may in fact be documents, passages, etc.
Within this general framework, we instantiate a specific design composed of three stages ($H_0$, $H_1$, and $H_2$), as shown in Figure~\ref{fig:overview}.

In our approach, each stage is unconstrained in its implementation other than the input--output specifications outlined above.
For example, a pipeline stage is not obligated to consider all candidates provided to it, and in fact, latency introduced by each stage can be controlled by truncating the number of input candidates.
Furthermore, each stage can choose to pay attention or ignore scores of the candidates it receives; in the latter case, the ranked list devolves into a set of unranked candidates.
In our experiments, we explore the latency--quality tradeoff space that is induced by this design flexibility (see Section~\ref{sec:results}).

\subsection{$H_0$: ``Bag of Words'' BM25}

The first stage $H_0$ receives as input the user query $q$ and produces top-$k_0$ candidates $R_0$.
In our implementation, the query is treated as a ``bag of words'' for ranking documents from the corpus using a standard inverted index based on the BM25 scoring function~\cite{robertson1995okapi}.
We use the Anserini IR toolkit~\cite{Yang_etal_SIGIR2017,Yang_etal_JDIQ2018},\footnote{\url{http://anserini.io/}} which is built on the popular open-source Lucene search engine.

BM25 is based on exact term matches, and all candidates must contain at least one term from the user's query.
However, since later BERT stages operate in continuous vector spaces, they have the ability to identify relevant candidates that do not have many matching terms.
Thus, it is critical in $H_0$ to optimize for recall to provide subsequent stages a diverse set of documents to work with.
On the other hand, precision is less of a concern because non-relevant documents can be discarded by later stages.

\subsection{$H_1$: monoBERT}

In general, the task of a re-ranking stage $H_n$ is to estimate a score $s_i$ quantifying how relevant a candidate $d_i \in R_{n-1}$ is to a query $q$.
Naturally, we expect that the ranking induced by these scores yields a higher metric (e.g., MAP or MRR) than the scores from the previous stage.

In stage $H_1$, we call monoBERT our pointwise re-ranker, which is a BERT model used as a binary relevance classifier.
Using the same notation as~\citet{devlin2018bert}, we feed the query $q$ as sentence A and the text of candidate $d_i$ as sentence B. We truncate the query to have at most 64 tokens.
We also truncate the candidate text such that the concatenation of query, candidate, and separator tokens have a maximum length of 512 tokens.
Given these limits, we observe that none of the queries or documents of the datasets used in our experiments (TREC CAR and MS MARCO) have to be truncated.

Once the segment is passed through the model, we use the $\left[\text{CLS}\right]$ vector as input to a single layer neural network to obtain a probability $s_i$ of the candidate $d_i$ being relevant to $q$.
We obtain a score $s_i$ for each candidate independently and generate a new list of candidates $R_1$ by keeping only the top-$k_1$ candidates based on these scores.

We train the model for re-ranking using cross-entropy loss: 
\begin{equation} 
\label{eq:monobert_loss}
L_\text{mono} = -\sum_{j \in J_{\text{pos}}} \log (s_j) - \sum_{j \in J_{\text{neg}}} \log (1 - s_j),
\end{equation}
where $J_{\text{pos}}$ is the set of indexes of the relevant candidates and $J_{\text{neg}}$ is the set of indexes of the non-relevant candidates in $R_0$. 

\subsection{$H_2$: duoBERT}
\label{sec:duoBERT_method}

The output $R_1$ from the previous stage is used as input to the pairwise re-ranker we call duoBERT.
Within the framework of ``learning to rank'', duoBERT can be characterized as a ``pairwise'' approach, while monoBERT can be characterized as a ``pointwise'' approach~\cite{LiuTY_FnTIR2009}. In this pairwise approach, the re-ranker estimates the probability $p_{i,j}$ of the candidate $d_i$ being more relevant than $d_j$.

This re-ranker is also a BERT model that takes as input the query as sentence A, candidate $d_i$ as sentence B, and candidate $d_j$ as sentence C. Similar to the original implementation, each sentence type (A, B, and C) has its own embedding that is summed to the token and positional embeddings.
We truncate the query, candidates $d_i$ and $d_j$ to 62, 223, and 223 tokens, respectively, so the entire sequence will have at most 512 tokens when concatenated with the $[\text{CLS}]$ token and the three separator tokens. 
Using the above truncation limits, in the datasets used in this work none of the queries are truncated, and less than 1\% of the documents are truncated.

We use the $[\text{CLS}]$ vector as input to a single layer neural network to obtain the probability $p_{i,j}$. Since there are $k_1$ candidates, $k_1 (k_1 - 1)$ probabilities are computed.
We then train the model with the following loss:
\begin{equation}
\label{eq:duobert_loss}
\begin{split}
L_\text{duo} = -&\sum_{i \in J_{\text{pos}}, j \in J_{\text{neg}}} \log (p_{i,j}) \\
&- \sum_{i \in J_{\text{neg}}, j \in J_{\text{pos}}} \log (1 - p_{i,j}),
\end{split}
\end{equation}
Note in the equation above that candidates $d_i$ and $d_j$ are never both relevant or not relevant.

At inference time, we aggregate the pairwise scores $p_{i,j}$ so that each document receives a single score $s_i$.
We investigate five different aggregation methods (\textsc{Sum}, \textsc{Binary}, \textsc{Min}, \textsc{Max}, and \textsc{Sample}):
\begin{equation}
\label{eq:pairwise_sum}
\textsc{Sum}: s_i = \sum_{j \in J_i} p_{i,j},
\end{equation}

\begin{equation}
\label{eq:pairwise_binary}
\textsc{Binary}: s_i = \sum_{j \in J_i} \mathbbm{1}_{p_{i,j} > 0.5},
\end{equation}

\begin{equation}
\label{eq:pairwise_min}
\textsc{Min}: s_i = \min_{j \in J_i} p_{i,j},
\end{equation}

\begin{equation}
\label{eq:pairwise_max}
\textsc{Max}: s_i = \max_{j \in J_i} p_{i,j},
\end{equation}

\begin{equation}
\label{eq:pairwise_sample}
\textsc{Sample}: s_i = \sum_{j \in J_i(m)} p_{i,j},
\end{equation}
where $J_i=\{0 \leq j < |R_1|, j \neq i\}$ and $m$ is the number of samples drawn without replacement from the set $J_i$.

The \textsc{Sum} method measures the pairwise agreement that candidate $d_i$ is more relevant than the rest of the candidates ${\{d_j\}}_{j \neq i}$.
The \textsc{Binary} method is inspired by the Condorcet method~\cite{montague2002condorcet}, which is a strong aggregation baseline~\cite{cormack2009reciprocal}.
The \textsc{Min} (\textsc{Max}) method measures the relevance of $d_i$ only against its strongest (weakest) competitor.
The \textsc{Sample} method aims to decrease the high inference costs of pairwise computations via sampling.

The final list of candidates $R_2$ is obtained by re-ranking the candidates in $R_1$ according to their scores ${s_i}$.
In our current design, the output $R_2$ is provided for human consumption, and serves as the input to computing the final evaluation metrics (e.g., MAP).

%$X = \{x | x \sim \mathcal{U}(J)\}$, where $$\mathcal{U}(J)$ is a uniform distribution over all the items in $J$.

\section{Experimental Setup}

A fortunate confluence of events has enabled the multi-stage ranking architecture we propose in this paper.
First, of course, is the innovation captured in BERT, as the latest refinement in a long stream of neural models that make heavy use of pre-training.
Second, and just as important, is the availability of data.
For document retrieval, most IR researchers have not had access to sufficient training data until recently.

As demonstrated by~\citet{lin2019neural}, in a limited data regime, it is not entirely clear that neural techniques actually perform better than well-tuned ``classic'' IR techniques; subsequent work by~\citet{Yang_etal_arXiv2019hype} show that the gains are modest at best.
Until recently, research in neural ranking models mostly took advantage of proprietary datasets derived from user behavior logs (which large organizations can gather in abundance).
Since these datasets cannot be shared, only a small set of researchers could productively work on neural ranking models and different models could not be easily compared; the combination of both factors hamper rapid progress.

Fortunately, the field has seen the release of two large-scale datasets for powering data-hungry neural models:\ MS MARCO~\cite{nguyen2016ms} and TREC CAR~\cite{dietz2017trec}.
We take advantage of both datasets to train our models, which we detail below.

\subsection{MS MARCO}

The Microsoft MAchine Reading COmprehension dataset (MS MARCO) is a large-scale resource created from approximately half a million anonymized questions sampled from Bing's search query logs.
We focus on the passage ranking task, where given a corpus of 8.8M passages extracted from 3.6M web documents, the system's goal is to retrieve passages that answer the question.
Each passage contains an average of 55 words (or 340 characters), and hence is relatively short---however, in order to maintain consistent terminology throughout this paper, we refer to these basic units of retrieval as ``documents.''

The training set (for the passage ranking task) comprises approximately 500k pairs of query and relevant document, and another 400M pairs of query and non-relevant documents.
The relevance judgments are provided by humans.
The development set contains 6,980 queries, with, on average, one relevant document per query.
Thus, a noteworthy property of this dataset is the sparsity of relevance judgments---as opposed to typical TREC test collections built using pooling~\cite{Voorhees_CLEF2002}, which have far fewer topics (usually around 50) but many more judgments per topic (typically, hundreds).
A blind, held-out evaluation set with 6,837 queries is also available, but without relevance judgments.
Evaluation on these queries is provided by the Microsoft organizers upon submission to the online leaderboard.
The official metric for this dataset is MRR@10.

\paragraph{Target Corpus Pre-training (TCP).}

Before training our models on the re-ranking task, we apply a two-phase pre-training. In the first phase, the model is pre-trained using Wikipedia (2.5B words) and the Toronto Book corpus (0.8B words) for one million iterations, as described by~\citet{devlin2018bert}.
In the second phase, we further pre-train the model on the MS MARCO corpus (0.5B words) for 100k iterations with a maximum sequence length of 512 tokens, batches of size 128, and learning rate of $5 \times 10^{-5}$. 
This second pre-training phase takes approximately 24 hours on a TPU v3.\footnote{
\url{https://cloud.google.com/tpu/}
}

\paragraph{Training.}

We fine-tune both monoBERT and duoBERT using a TPU v3 with a batch size of 128 (128 sequences $\times$ 512 tokens = 16,384 tokens/batch) for 100k iterations, which takes approximately 24 hours.
This corresponds to training on 12.8M (100k $\times$ 128) query--document pairs.
We could not see any improvement on the dev set when training for another three days, which is equivalent to seeing 50M query--document pairs in total.
To avoid biasing our model towards predicting non-relevant labels, which are approximately 1000 times more frequent in the training set, we build each batch by sampling an equal amount of relevant and non-relevant passages.

For both models, we use Adam~\citep{kingma2014adam} with the initial learning rate set to $3 \times 10^{-6}$, $\beta_1 = 0.9$, $\beta_2 = 0.999$, L2 weight decay of 0.01, learning rate warmup over the first 10,000 steps, and linear decay of the learning rate.
Dropout probability is set to $0.1$ in all layers.

\paragraph{Inference.} 

In our base configuration, we use top-$k_0=1000$ and top-$k_1=50$ candidates as input to monoBERT and duoBERT, respectively.
Our experiments, however, include ablation settings as well as different parameterizations to characterize the contributions of each component as well as the latency--quality tradeoff space.

\subsection{TREC CAR}

Our second dataset is from the Complex Answer Retrieval (CAR) Track at the 2017 Text Retrieval Conference (TREC), whose aim is to explore passage-level retrieval techniques for simple fact and entity-centric needs~\cite{dietz2017trec}.
The primary dataset is synthetically constructed by exploiting the hierarchical structure of Wikipedia:\ ``queries'' are constructed by concatenating a Wikipedia article title with the title of one of its sections.
The relevant documents are the paragraphs within that section.
The corpus consists of cleaned paragraphs from English Wikipedia, except for the abstracts, totaling 29M documents, with an average of 60 words (or 380 characters) per document.
The released dataset has five predefined folds, and we use the first four as a training set (approximately 3M queries), and the remaining as a validation set (approximately 700k queries). The test set is the same one used to evaluate the submissions to TREC 2017 CAR (2,254 queries).

Although the TREC 2017 CAR organizers provide manual annotations for the test set, only the top five documents retrieved by systems that submitted to the official evaluation have manual annotations.
The sparsity of these judgments means that it is difficult to fairly evaluate runs that did not participate in the original evaluation.
Hence, in this paper we evaluate using the automatic annotations, which provide a richer set of judgments.
Per the official TREC CAR evaluation, we use Mean Average Precision (MAP) as the evaluation metric.

\paragraph{Training.}

Both monoBERT and duoBERT are trained in the same manner as for MS MARCO, with the same hyperparameter settings. However, there is an important difference. The official pre-trained BERT models\footnote{
\url{https://github.com/google-research/bert}
} 
are pre-trained on the full Wikipedia, and therefore they have seen, although in an unsupervised way, Wikipedia documents that are used in the test set of TREC CAR. Thus, to avoid this leak of test data into training, we pre-train the BERT re-ranker only on the half of Wikipedia used by TREC CAR's training set, which contains 1.1B words.

For fine-tuning, we generate our query--document pairs by retrieving the top ten documents from the entire TREC CAR corpus using BM25.
This means that we end up with 30M example pairs (3M queries $\times$ 10 candidates/query) to train our model.
We train it for 100k iterations, or 12.8M examples (100k iterations $\times$ 128 pairs/batch).
Similar to the MS MARCO experiments, we did not see any gain on the dev set by training the model longer.

\section{Results}
\label{sec:results}

\begin{table}[t]
%\centering\resizebox{\columnwidth}{!}{
\begin{small}
\begin{tabular}{l|ll}
\noalign{\vskip 1mm}
\toprule
\noalign{\vskip 1mm}
{\bf Method} & Dev & Eval \\
\noalign{\vskip 1mm}
\toprule
\noalign{\vskip 1mm}
BM25 (Microsoft Baseline) & 16.7 & 16.5 \\
%KNRM~\citep{xiong2017end} & 21.8 & 19.8 \\
%Conv-KNRM~\citep{dai2018convolutional} & 29.0 & 27.1 \\
IRNet & 27.8 &  28.1 \\
monoBERT (Jan 2019) & 36.5 & 35.9 \\
\noalign{\vskip 1mm}
\midrule
\noalign{\vskip 1mm}
Anserini (BM25) & 18.7 & 19.0\\
+ monoBERT & 37.2 & 36.5\\
+ monoBERT + duoBERT$_\textsc{Max}$ & 32.6 & -\\
+ monoBERT + duoBERT$_\textsc{Min}$ & 37.9 & -\\
+ monoBERT + duoBERT$_\textsc{Sum}$ & 38.2 & 37.0\\
+ monoBERT + duoBERT$_\textsc{Binary}$ & 38.3 & -\\
+ monoBERT + duoBERT$_\textsc{Sum}$ + TCP & 39.0 & 37.9\\
\noalign{\vskip 1mm}
\midrule
\noalign{\vskip 1mm}
Leaderboard best & 39.7 & 38.3 \\
\noalign{\vskip 1mm}
\toprule
\end{tabular}
\end{small}
%}
\caption{MS MARCO Results.} 
\label{tab:ms-marco}
\end{table}

Results on the MS MARCO dataset are shown in Table~\ref{tab:ms-marco}.
The first row shows the BM25 baseline provided by Microsoft.
Our initial application of BERT to the MS MARCO dataset, denoted by the entry monoBERT (Jan 2019), was published in January~2019~\citep{nogueira2019passage}.
On the evaluation data, it surpassed the previous best entry IRNet (submitted just five days earlier) by nearly eight points.
This entry implements what we refer to as monoBERT here, albeit with a few minor differences, explained below.
We are, based on official leaderboard records, the first to adapt BERT to the MS MARCO dataset, and to our knowledge, our model represents the first application of BERT to any retrieval task.
We further note that {\it every} subsequent submission on the MS MARCO leaderboard (as of October 2019) exploits BERT in some capacity (evidenced by ``BERT'' appearing in every submission name).
Given the availability of our source code on GitHub, it is likely that many of these entries are derived from or build on monoBERT, or are at least inspired by our innovation.\footnote{Unfortunately, we cannot know with absolute certainty because most of the submissions are not paired with associated papers and source code.}

Our BM25 baseline with Anserini is shown in the first row of the second block of Table~\ref{tab:ms-marco}; in our multi-stage ranking architecture, this is $R_0$, the output of $H_0$.
Although both runs purport to implement BM25, Anserini is two points better than the Microsoft baseline.
Our recall at 1000 hits is 85.7\%, compared to only 81.5\% from Microsoft's implementation.
It is a well-known fact in IR that different systems implementing the same scoring function might report very different results~\cite{Muhleisen_etal_SIGIR2014,Lin_etal_ECIR2016}, owing to details such as tokenization, stopword selection, stemming, and parameter tuning.
Thus, the differences between Anserini and the Microsoft baseline are not surprising.

By applying the monoBERT stage $H_1$ to the top 1000 ranked list from Anserini ($H_0$ with $k_0 = 1000$), we observe a gain of 17.5 points.
This result is slightly better than the monoBERT entry from January 2019 because that submission re-ranked the Microsoft baseline (a slightly worse $H_0$, in essence).
Other minor differences include a refactored codebase to improve reusability and readability.

Adding the duoBERT stage $H_2$ with the \textsc{Sum} aggregation method (Equation~\ref{eq:pairwise_sum}), denoted duoBERT$_\textsc{Sum}$, improves over monoBERT alone by 0.5 points on the held-out evaluation set.
In this setting, duoBERT considers the top 50 candidates from $H_1$, and thus requires an additional $50\times49$ BERT inferences to compute the final ranking (the time required for aggregation is negligible).
This improvement in MRR, of course, comes at a cost in increased latency, an issue we explore in more detail below.
The entry marked duoBERT$_\textsc{Max}$ shows that the \textsc{Max} aggregation method (Equation~\ref{eq:pairwise_max}) performs quite poorly, and in fact makes monoBERT results worse. We find that the \textsc{Binary} method (Equation~\ref{eq:pairwise_binary}) performs slightly better (0.1 points) than \textsc{Sum} on the development set.
Given these results, we abandon the \textsc{Max} aggregation method in subsequent experiments.

Note that official figures from the held-out evaluation set are not available for all conditions because obtaining those values requires formal submission of runs to the MS MARCO organizers. 
As good experimental practice, in order to avoid too much ``unnecessary probing'' of the held-out test data, we only submitted what we felt to be the most promising conditions.

Finally, pre-training on the target corpus (monoBERT + duoBERT$_\textsc{Sum}$ + TCP) improves MRR@10 by another 0.8 points.
This result is in line with recent work that shows improvements with target corpus pre-training over out-of-domain corpus pre-training~\cite{beltagy2019scibert,raffel2019exploring}.

\begin{table}[t]
%\vskip 0.05in
%\centering\resizebox{\columnwidth}{!}{
\centering
\begin{small}
\begin{tabular}{l|l}
\noalign{\vskip 1mm}
\toprule
\noalign{\vskip 1mm}
{\bf Method} & MAP\\
\noalign{\vskip 1mm}
\toprule
\noalign{\vskip 1mm}
BM25~\cite{kashyapi2018trema} & 13.0 \\
Co-PACRR~\citep{macavaney2017contextualized} & 14.8\\
\noalign{\vskip 1mm}
\midrule
\noalign{\vskip 1mm}
BM25 (Anserini) & 15.3 \\
+ monoBERT & 34.8\\
+ monoBERT + duoBERT$_\textsc{Max}$ & 32.6 \\
+ monoBERT + duoBERT$_\textsc{Sum}$ & 36.9 \\
+ monoBERT + duoBERT$_\textsc{Binary}$ & 36.9 \\
\noalign{\vskip 1mm}
\toprule
\end{tabular}
\end{small}
%}
\caption{Main Result on TREC 2017 CAR.} 
\label{tab:car}
\end{table}

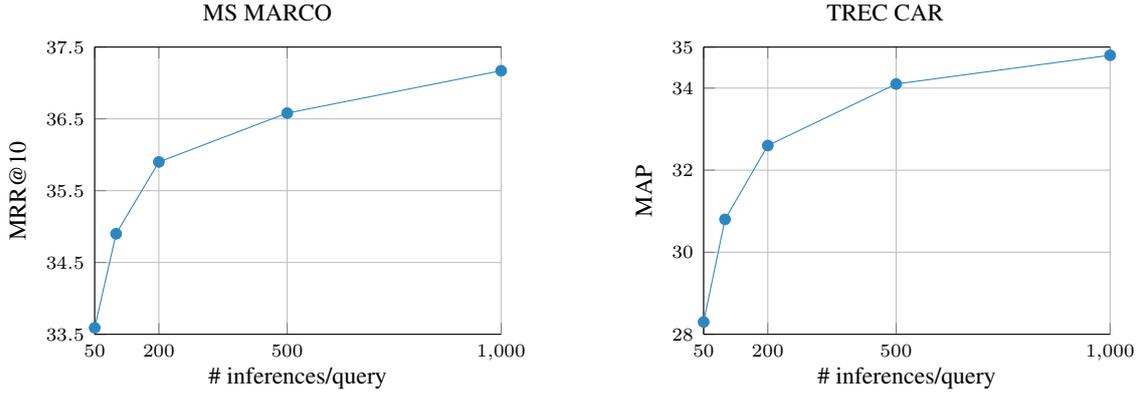
\begin{figure*}[t]
\centering
\begin{tikzpicture}[scale = 1.0]
\begin{axis}[
width=0.9\columnwidth,
height=0.70\columnwidth,
legend cell align=left,
legend style={font=\tiny},
mark options={mark size=3},
font=\scriptsize,
axis y line*=left,
xmin=50, xmax=1000,
ymin=33.5, ymax=37.5,
log ticks with fixed point,
xtick={50, 200, 500, 1000},
ytick={33.5, 34.5, 35.5, 36.5, 37.5},
legend pos=south east,
xmajorgrids=true,
ymajorgrids=true,
xlabel style={font = \small, yshift=1ex},
xlabel=\# inferences/query,
ylabel=MRR@10,
ylabel style={font = \small, yshift=0ex}]
    \addplot[mark=*,g-blue, mark options={scale=1}] plot coordinates {
    (50, 33.59)(100, 34.90)(200, 35.90)(500, 36.58)(1000, 37.17)
    };
\end{axis}
\node[above, font=\small] at (current bounding box.north) {MS MARCO};
\end{tikzpicture}
\hspace{1.0cm}
\begin{tikzpicture}[scale = 1.0]
%\centering
\begin{axis}[
width=0.9\columnwidth,
height=0.70\columnwidth,
legend cell align=left,
legend style={font=\tiny},
mark options={mark size=3},
font=\scriptsize,
axis y line*=left,
xmin=50, xmax=1000,
ymin=28.0, ymax=35.0,
log ticks with fixed point,
xtick={50, 200, 500, 1000},
ytick={28.0, 30.0, 32.0, 34.0, 35.0},
legend pos=south east,
xmajorgrids=true,
ymajorgrids=true,
xlabel style={font = \small, yshift=1ex},
xlabel=\# inferences/query,
ylabel=MAP,
ylabel style={font = \small, yshift=0ex}]
    \addplot[mark=*,g-blue, mark options={scale=1}] plot coordinates {
    (50, 28.30)(100, 30.8)(200, 32.6)(500, 34.1)(1000, 34.8)
    };
\end{axis}
\node[above, font=\small] at (current bounding box.north) {TREC CAR};
\end{tikzpicture}

\vspace{-2mm}
\caption{Number of inferences per query vs.\ effectiveness on the MS MARCO and the TREC CAR datasets when varying the number of candidates $k_0$ fed to monoBERT.}
\label{fig:monobert_cutoff}
\end{figure*}

\begin{figure*}[t]
\centering
\begin{tikzpicture}[scale = 1.0]
\begin{axis}[
width=0.90\columnwidth,
height=0.95\columnwidth,
legend cell align=left,
mark options={mark size=3},
font=\scriptsize,
axis y line*=left,
xmin=0, xmax=2450,
ymin=37.0, ymax=38.5,
log ticks with fixed point,
xtick={0, 380, 870, 1560, 2450},
ytick={37.0, 37.5, 38.0, 38.5},
legend pos=south east,
xmajorgrids=true,
ymajorgrids=true,
xlabel style={font = \small, yshift=1ex},
xlabel=\# inferences/query,
ylabel= MRR@10,
ylabel style={font = \small, yshift=0ex}]
    \addplot[mark=*,g-blue, mark options={scale=1}] plot coordinates {
    (0, 37.2)(90, 37.95)(380, 38.12)(870, 38.14)(1560, 38.15)(2450, 38.19)
    };
    \addlegendentry{\textsc{Sum}}
    \addplot[mark=triangle,red, mark options={scale=1}] plot coordinates {
    (0, 37.2)(90, 37.45)(380, 38.27)(870, 38.3)(1560, 38.31)(2450, 38.31)
    };
    \addlegendentry{\textsc{Binary}}
    \addplot[mark=o,g-purple, mark options={scale=1}] plot coordinates {
    (0, 37.2)(90, 37.81)(380, 37.87)(870, 37.87)(1560, 37.88)(2450, 37.89)
    };
    \addlegendentry{\textsc{Min}}
    %\addplot[mark=x,g-purple, mark options={scale=1}] plot coordinates {
    %(0, 37.2)(190, 36.97)(380, 37.04)(570, 37.15)(760, 37.17)(950, 37.2)
    %};
    %\addlegendentry{\textsc{Sample}, m=10}
    \addplot[line width=1pt, dotted, mark=*,gray, mark options={scale=1}] plot coordinates {
    (0, 37.2)(190, 37.5)(380, 37.52)(570, 37.61)(760, 37.7)(950, 37.8)
    };
    \addlegendentry{\textsc{Sample}, m=20}
    %\addplot[mark=x,green, mark options={scale=1}] plot coordinates {
    %(0, 37.2)(290, 37.77)(580, 37.89)(870, 37.95)(1160, 37.99)(1470, 38.03)
    %};
    %\addlegendentry{\textsc{Sample}, m=30}
    \addplot[line width=1pt, dashed, mark=*,gray, mark options={scale=1}] plot coordinates {
    (0, 37.2)(390, 37.97)(780, 38.03)(870, 38.05)(1170, 38.12)(1560, 38.13)
    };
    \addlegendentry{\textsc{Sample}, m=40}
\end{axis}
\node[above, font=\small] at (current bounding box.north) {MS MARCO};
\end{tikzpicture}
\hspace{1.0cm}
\begin{tikzpicture}[scale = 1.0]
%\centering
\begin{axis}[
width=0.90\columnwidth,
height=0.95\columnwidth,
legend cell align=left,
mark options={mark size=3},
font=\scriptsize,
axis y line*=left,
xmin=0, xmax=2450,
ymin=34.5, ymax=37.0,
log ticks with fixed point,
xtick={0, 380, 870, 1560, 2450},
ytick={34.5, 35.0, 35.5, 36.0, 36.5, 37.0},
legend pos=south east,
xmajorgrids=true,
ymajorgrids=true,
xlabel style={font = \small, yshift=1ex},
xlabel=\# inferences/query,
ylabel=MAP,
ylabel style={font = \small, yshift=0ex}]
    \addplot[mark=*,g-blue, mark options={scale=1}] plot coordinates {
    (0, 34.8)(90, 35.47)(380, 36.69)(870, 36.80)(1560, 36.92)(2450, 36.93)
    };
    \addlegendentry{\textsc{Sum}}
    \addplot[mark=triangle,red, mark options={scale=1}] plot coordinates {
    (0, 34.8)(90, 35.54)(380, 36.35)(870, 36.69)(1560, 36.90)(2450, 36.90)
    };
    \addlegendentry{\textsc{Binary}}
    \addplot[mark=o,g-purple, mark options={scale=1}] plot coordinates {
    (0, 34.8)(90, 35.52)(380, 36.21)(870, 36.27)(1560, 36.38)(2450, 36.42)
    };
    \addlegendentry{\textsc{Min}}
    %\addplot[mark=x,g-purple, mark options={scale=1}] plot coordinates {
    %(0, 34.8)(90, 34.88)(180, 35.39)(270, 35.45)(360, 35.45)(450, 35.57)
    %};
    %\addlegendentry{\textsc{Sample}, m=10}
    \addplot[line width=1pt, dotted, mark=*,gray, mark options={scale=1}] plot coordinates {
    (0, 34.8)(190, 35.83)(380, 36.38)(570, 36.45)(760, 36.47)(950, 36.54)
    };
    \addlegendentry{\textsc{Sample}, m=20}
    %\addplot[mark=x,green, mark options={scale=1}] plot coordinates {
    %(0, 34.8)(290, 35.29)(580, 36.37)(870, 36.55)(1160, 36.61)(1470, 36.61)
    %};
    %\addlegendentry{\textsc{Sample}, m=30}
    \addplot[line width=1pt, dashed, mark=*,gray, mark options={scale=1}] plot coordinates {
    (0, 34.8)(390, 35.77)(780, 36.34)(870, 36.41)(1170, 36.49)(1560, 36.71)
    };
    \addlegendentry{\textsc{Sample}, m=40}
\end{axis}
\node[above, font=\small] at (current bounding box.north) {TREC CAR};
\end{tikzpicture}

\vspace{-2mm}
\caption{Number of inferences per query vs.\ the effectiveness of duoBERT when varying the number of candidates $k_1$. Each curve has six points that correspond to $k_1=\{0, 10, 20, 30, 40, 50\}$, where $k_1=0$ corresponds to monoBERT. The values in the {\it x}-axis are computed as $k_1 \times (k_1 - 1)$ for \textsc{Sum}, \textsc{Binary}, and \textsc{Min}, and $k_1 \times (m - 1)$ for \textsc{Sample}. To avoid clutter, plots for \textsc{Sample} at $m = \{10, 30\}$ are omitted.}
\label{fig:duobert_cutoff}
\end{figure*}
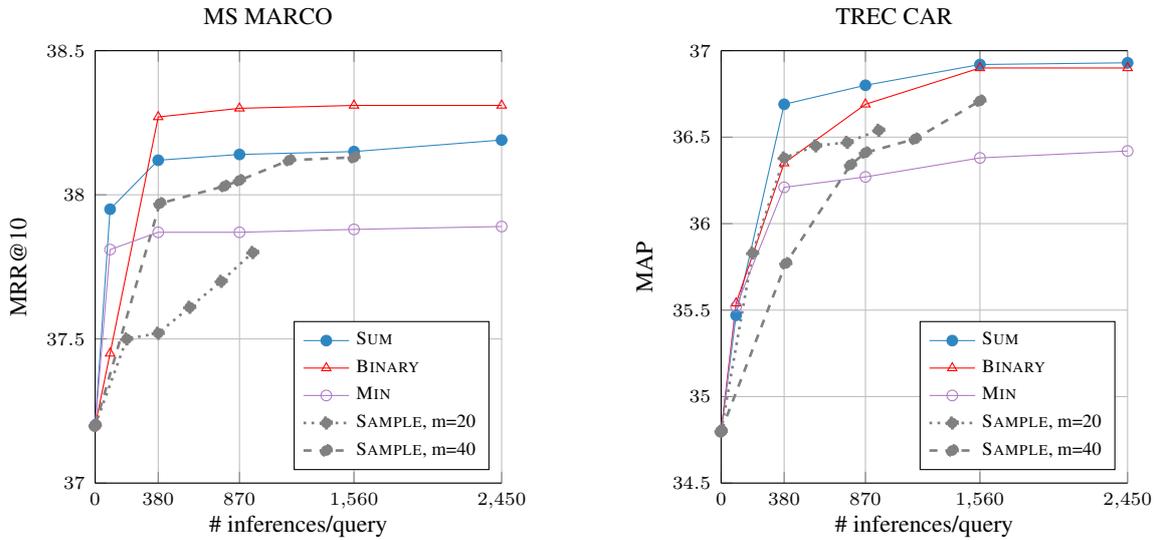

Results for TREC CAR are presented in Table~\ref{tab:car}, organized in a similar manner as Table~\ref{tab:ms-marco}.
We see similar trends on this dataset.
Again, Anserini's implementation of BM25 leads to 2.3 MAP points improvement over another Lucene-based implementation from~\citet{kashyapi2018trema}.
It is also 0.5 MAP points higher than the best entry from TREC 2017 CAR~\cite{macavaney2017contextualized}.
The monoBERT model gives an impressive jump of 19.5 MAP points over the BM25 baseline and duoBERT$_\textsc{Sum}$ or duoBERT$_\textsc{Binary}$ provides another improvement of 2.1 points. 
To our knowledge, this is the best-known result on this dataset.
Note that we do not report target corpus pre-training results on TREC CAR because its target corpus is the same as the original BERT pre-training corpus, i.e., English Wikipedia.

In general, we notice that improvements from our BERT models are larger on TREC CAR than they are on MS MARCO.
We believe this is primarily due to the evaluation metric:\ improvements in MRR@10 are much harder to achieve, since only the first correct answer contributes to the score, while better rankings of {\it all} relevant documents improve the MAP score.
Additionally, MRR@10 is a highly discrete metric (there are only 11 possible values), and these values are arranged such that large gains in effectiveness are only possible in the early ranks (thus increasing the level of task difficulty).

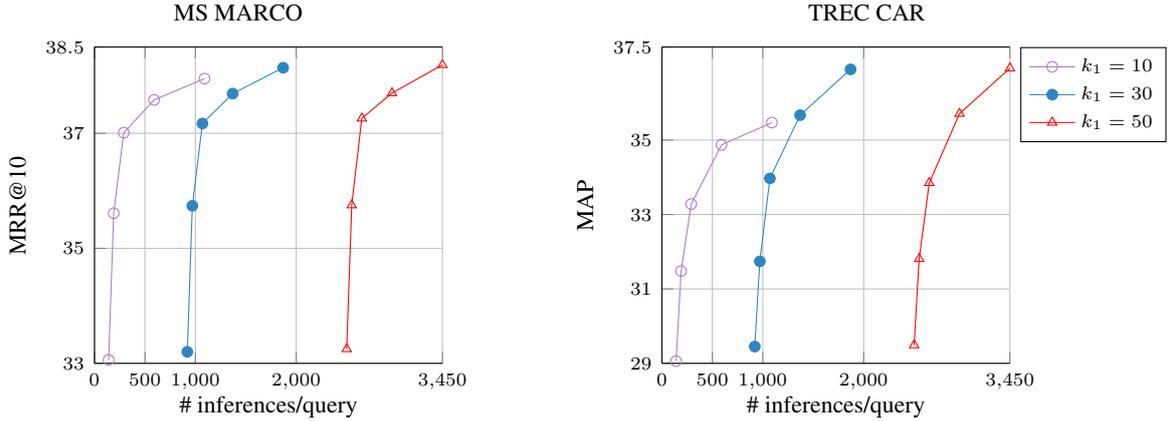
\begin{figure*}[t]
\centering
\begin{tikzpicture}[scale = 1.0]
\begin{axis}[
width=0.80\columnwidth,
height=0.75\columnwidth,
legend cell align=left,
mark options={mark size=3},
font=\scriptsize,
axis y line*=left,
xmin=0, xmax=3450,
ymin=33.0, ymax=38.5,
log ticks with fixed point,
xtick={0, 500, 1000, 2000, 3450},
ytick={33.0, 35.0, 37.0, 38.5},
legend pos=north west,
xmajorgrids=true,
ymajorgrids=true,
xlabel style={font = \small, yshift=1ex},
xlabel=\# inferences/query,
ylabel= MRR@10,
ylabel style={font = \small, yshift=0ex}]
    \addplot[mark=o,g-purple, mark options={scale=1}] plot coordinates {
    (140, 33.06)(190, 35.61)(290, 37.01)(590, 37.58)(1090, 37.95)
    };
    %\addlegendentry{$k_1=10$}
    \addplot[mark=*,g-blue, mark options={scale=1}] plot coordinates {
    (920, 33.20)(970, 35.74)(1070, 37.17)(1370, 37.69)(1870, 38.14)
    };
    %\addlegendentry{$k_1=30$}
    \addplot[mark=triangle,red, mark options={scale=1}] plot coordinates {
    (2500, 33.25)(2550, 35.75)(2650, 37.26)(2950, 37.70)(3450, 38.19)
    };
    %\addlegendentry{$k_1=50$}
\end{axis}
\node[above, font=\small] at (current bounding box.north) {MS MARCO};
\end{tikzpicture}
\hspace{1.0cm}
\begin{tikzpicture}[scale = 1.0]
%\centering
\begin{axis}[
width=0.80\columnwidth,
height=0.75\columnwidth,
legend cell align=left,
mark options={mark size=3},
font=\scriptsize,
axis y line*=left,
xmin=0, xmax=3450,
ymin=29.0, ymax=37.5,
log ticks with fixed point,
xtick={0, 500, 1000, 2000, 3450},
ytick={29.0, 31.0, 33.0, 35.0, 37.5},
legend pos=outer north east,
xmajorgrids=true,
ymajorgrids=true,
xlabel style={font = \small, yshift=1ex},
xlabel=\# inferences/query,
ylabel=MAP,
ylabel style={font = \small, yshift=0ex}]
    \addplot[mark=o,g-purple, mark options={scale=1}] plot coordinates {
    (140, 29.06)(190, 31.48)(290, 33.28)(590, 34.87)(1090, 35.47)
    };
    \addlegendentry{$k_1=10$}
    \addplot[mark=*,g-blue, mark options={scale=1}] plot coordinates {
    (920, 29.45)(970, 31.74)(1070, 33.97)(1370, 35.67)(1870, 36.9)
    };
    \addlegendentry{$k_1=30$}
    \addplot[mark=triangle,red, mark options={scale=1}] plot coordinates {
    (2500, 29.49)(2550, 31.81)(2650, 33.85)(2950, 35.71)(3450, 36.93)
    };
    \addlegendentry{$k_1=50$}
%\end{semilogxaxis}
\end{axis}
\node[above, font=\small] at (current bounding box.north) {TREC CAR};
\end{tikzpicture}

\vspace{-2mm}
\caption{Number of inferences per query vs.\ the effectiveness of duoBERT$_\textsc{Sum}$ when varying the number of candidates $k_0$ and $k_1$. Each curve has five points that correspond to $k_0=\{50, 100, 200, 500, 1000\}$. The number of inferences per query is calculated as $k_0 + k_1(k_1-1)$.}
\label{fig:duobert_k0k1}
\end{figure*}

\subsection{Tradeoffs with monoBERT}

The experimental results presented above capture monoBERT and duoBERT settings that focus on obtaining the best output quality.
Our next set of experiments explore different parameterizations of the multi-stage ranking architecture that realizes different quality--latency tradeoffs.

For monoBERT, the number of candidates $k_0$ is the control ``knob'':\ latency increases linearly as we consider more candidates, but effectiveness increases as well.
This relationship is shown in Figure~\ref{fig:monobert_cutoff} for MS MARCO on the left and TREC CAR on the right.
To aid in comparisons with duoBERT experiments below, the {\it x}-axis shows the number of inferences performed per query, which is exactly the same as $k_0$, since each query--candidate pair from $R_1$ serves as an input to monoBERT.

As expected, we see diminishing returns with larger $k_0$ values on both datasets.
For example, compared to $k_0 = 1000$, on both datasets we can achieve more than half the gain in effectiveness with only around a fifth of the number of inferences.
These curves also highlight the inadequacy of BM25 scores alone, since with a deep candidate list $R_0$, monoBERT is considering documents that have quite low BM25 scores.

\subsection{Tradeoffs with duoBERT}

Similar to monoBERT, we can control the latency--quality tradeoff of duoBERT by considering different $k_1$ values.
In these experiments, $k_0$ is fixed at 1000, and in 
Figure~\ref{fig:duobert_cutoff} we plot changes in effectiveness (MRR@10 for MS MARCO on the left, MAP for TREC CAR on the right) as a function of latency (inferences/query) for different values of $k_1$.
We find that actual inference latencies (measured in milliseconds) for duoBERT and monoBERT are comparable, and so the number of inferences per query provides a natural abstract time unit to support meaningful comparisons.

In the figure, each curve represents an aggregation technique and contains six points that correspond to $k_1=\{0, 10, 20, 30, 40, 50\}$.
The leftmost point, $k_1=0$, corresponds to monoBERT only, which allows us to quantify the additive impact of the duoBERT stage on top monoBERT results.
The values for the \textsc{Sample} method represent the average of ten trials.

Of the four aggregation methods compared, \textsc{Binary} yields the highest effectiveness on the MS MARCO dataset, albeit by a small margin over \textsc{Sum}.
On TREC CAR, \textsc{Binary} and \textsc{Sum} are very close, although \textsc{Sum} appears to be slightly better, especially at lower cutoffs.
The \textsc{Sample} method has a lower effectiveness than \textsc{Binary} and \textsc{Sum} for any fixed number of inferences per query.
This result shows that the top-$k_1$ candidates from monoBERT are a closer approximation of the true relevance ranking than uniformly sampling from a larger candidate set.
This is an interesting result:\ given the choice of sampling from a larger candidate set or exhaustively enumerating all pairs from a smaller candidate set, the latter option always seems to yield better answers.

Considering these results, it seems that a good operating point is $k_1=20$ with \textsc{Binary} aggregation on MS MARCO and \textsc{Sum} aggregation on TREC CAR.
In both cases, we obtain close to the maximum achievable score, with only a 40\% increase in latency compared to monoBERT only (whereas $k_1=50$, \textsc{Sum} or \textsc{Binary}, more than doubles the number of inferences required over monoBERT).

\begin{table*}[ht]
	\scriptsize
	\centering
	\begin{tabular}{|C{2cm}|C{7.8cm}|c|c|c|}
		\hline
		\multirow{2}{*}{\textbf{Query}} & \multirow{2}{*}{\textbf{Sample Passage}} & \multirow{2}{*}{\textbf{Label}} & \multicolumn{2}{c|}{\textbf{Rank}} \\ \cline{4-5} 
		& &  & \textbf{Baseline} & \textbf{Comparison} \\ \hline \hline
		    \multirow{2}{\hsize}{\textbf{who wrote song killing the blues}}
 & \attn{Killing The Blues}{99} by Robert Plant and Alison Krauss. This was written by Chris Isaak's bass guitarist Roly Salley, and was originally the title track of Salley's 2005 solo album. This \attn{song}{99} was used in an advertising campaign for the chain store JC Penney, which features sentimental images of heartland Americana, such as family reunions and Fourth of July celebrations. & R & BM25: 621
 & monoBERT: 1 \\ \cline{2-5} 
    & \attn{Who wrote}{99} the \attn{blues}{99} \attn{song}{99} Crossroads Cross Road \attn{Blues}{99} is one of Delta \attn{Blues}{99} singer Robert Johnson's most famous \attn{songs}{99}. Who \attn{wrote}{99} the \attn{song}{99} '\attn{Blue}{99} Shades.. Frank Ticheli wrote the \attn{song}{99} '\attn{Blue}{99} Shades'. It is a concert piece with allusions...
    & N & BM25: 1
 & monoBERT: 9 \\ \cline{1-1} \cline{2-5} 
    \multirow{2}{\hsize}{\textbf{what causes low liver enzymes}} &  \attn{Reduced}{50} production of \attn{liver enzymes}{99} may \attn{indicate}{50} dysfunction of the \attn{liver}{99}. This article explains the causes and symptoms of \attn{low liver enzymes}{99}. Scroll down to know how the production of the enzymes can be accelerated.
 & R &  monoBERT: 47
 & duoBERT: 1 \\  \cline{2-5} 
   & Other \attn{causes}{99} of \underline{elevated} \attn{liver enzymes}{99} may include: Alcoholic hepatitis (severe liver inflammation caused by excessive alcohol consumption) Autoimmune hepatitis (liver inflammation caused by an autoimmune disorder) Celiac disease (small intestine damage caused by gluten) Cytomegalovirus (CMV) infection.
    & N & monoBERT: 1
 & duoBERT: 7 \\ \cline{1-1} \cline{2-5} 
	\end{tabular}
	\caption{Comparison of BM25 vs.\ monoBERT, and monoBERT vs.\ duoBERT, showing result ranks of answers. (\textbf{N}:\ not relevant, \textbf{R}:\ relevant)}
	\label{tab:samples}
\end{table*}

\subsection{Multi-Stage Tradeoffs}

Our next set of experiments quantify the tradeoffs when changing both $k_0$ and $k_1$; results are shown in Figure~\ref{fig:duobert_k0k1}.
Since inference times are approximately the same between monoBERT and duoBERT, we can quantify latency by the number of inferences.

On both datasets, the most computationally expensive point in the blue curve ($k_0=1000$ and $k_1=50$) has a much higher effectiveness than the least expensive point in the red curve ($k_0=50$ and $k_1=50$).
This provides an example that analyzing multiple cutoffs jointly can improve our understanding of the tradeoff space.

\subsection{Qualitative Analyses}

Finally, we conduct qualitative analyses by sampling retrieved passages from three methods:\ BM25, monoBERT, and duoBERT${_\textsc{Sum}}$.
A few examples are shown in Table~\ref{tab:samples}. 
From the first two examples, we can see that BM25 tries to maximize unigram matches between queries and passages, and thus often neglects $n$-grams, while monoBERT learns to assign a high matching score to $n$-grams.
This also shows an example where a high BM25 score---that comes from repeated instances of query terms---can be misleading.
Our monoBERT model, at least in this example, does not appear to be fooled.

From the last two samples in Table~\ref{tab:samples}, we can see that duoBERT matches the synonyms between ``low'' in the query and ``reduced'' in the passage, while monoBERT fails to distinguish ``low'' in the query and ``elevated'' in the passage.

\section{Future Work and Conclusions}

While our work is firmly situated in the context of multi-stage ranking architectures, it makes sense to discuss the broader landscape of applying neural models to document ranking.
Search-related tasks, almost by definition, need to consider a large corpus, and thus it is impractical to apply inference over {\it all} documents for a given query.
This simple fact necessitates reliance on standard ``bag of words'' techniques to reduce the ``working set'' that is presented to neural models.

Such a design, however, is inelegant, which has led researchers to explore alternatives that are able to directly perform ranking.
The most promising approach is formulated as a representational learning problem, where the task is to learn some non-linear transformation of queries and documents (i.e., using a neural network) such that documents relevant to a query have high similarities in terms of a simple metric such as cosine similarity~\cite{Henderson:1705.00652:2017,Zamani:2018:NRN:3269206.3271800,Ji:2019:EIN:3308558.3313576}.
This, in essence, transforms neural ranking into approximate nearest-neighbor search once queries and documents have been mapped into the learned representational space.

While this is indeed a promising approach, and has seen production deployment in limited contexts~\cite{Henderson:1705.00652:2017}, this thread of research is better characterized as exploratory.
It is unclear whether representational learning is sufficient to boil the complex notion of relevance down to simple similarity computations---and if it isn't, the complete end-to-end retrieval architecture will need to involve multiple stages anyway.
In contrast, multi-stage ranking architectures are mature, well understood, easy to deploy, and proven in production.

Our future work aims to build the stages of the pipeline jointly, in which hyperparameters are automatically tuned for end-to-end performance. 
Also, explicitly using scoring signals from previous stages of the pipeline in later stages has the potential to increase overall effectiveness as more information is shared among stages.
Lastly, current BERT-based models can only handle documents that are a few sentences long (at the most).
Models that can handle longer documents without truncation, such as \citet{Yilmaz_etal_EMNLP2019}, should be evaluated on datasets such as the MS MARCO document ranking task.
Overall, we believe that multi-stage ranking architectures pave the way to {\it practical} deployment of complex and computationally-intensive neural models.

\section*{Acknowledgments}

RN and KC thank support by NVIDIA and CIFAR and were partly supported by Samsung Advanced Institute of Technology (Next Generation Deep Learning:\ from pattern recognition to AI) and Samsung Electronics (Improving Deep Learning using Latent Structure).
WY and JL thank support by the Natural Sciences and Engineering Research Council (NSERC) of Canada, with additional computational resources provided by Compute Ontario and Compute Canada. 

\bibliography{main}

\begin{thebibliography}{46}
\expandafter\ifx\csname natexlab\endcsname\relax\def\natexlab#1{#1}\fi

\bibitem[{Asadi and Lin(2013)}]{Asadi_Lin_SIGIR2013}
Nima Asadi and Jimmy Lin. 2013.
\newblock Effectiveness/efficiency tradeoffs for candidate generation in
  multi-stage retrieval architectures.
\newblock In \emph{Proceedings of the 36th Annual International ACM SIGIR
  Conference on Research and Development in Information Retrieval (SIGIR
  2013)}, pages 997--1000, Dublin, Ireland.

\bibitem[{Bajaj et~al.(2018)Bajaj, Campos, Craswell, Deng, Gao, Liu, Majumder,
  McNamara, Mitra, Nguyen, Rosenberg, Song, Stoica, Tiwary, and
  Wang}]{nguyen2016ms}
Payal Bajaj, Daniel Campos, Nick Craswell, Li~Deng, Jianfeng Gao, Xiaodong Liu,
  Rangan Majumder, Andrew McNamara, Bhaskar Mitra, Tri Nguyen, Mir Rosenberg,
  Xia Song, Alina Stoica, Saurabh Tiwary, and Tong Wang. 2018.
\newblock {MS} {MARCO}: {A} human generated {MAchine} {Reading} {COmprehension}
  dataset.
\newblock \emph{arXiv:1611.09268v3}.

\bibitem[{Beltagy et~al.(2019)Beltagy, Cohan, and Lo}]{beltagy2019scibert}
Iz~Beltagy, Arman Cohan, and Kyle Lo. 2019.
\newblock {Sci\-BERT}: Pretrained contextualized embeddings for scientific
  text.
\newblock \emph{arXiv:1903.10676}.

\bibitem[{Cambazoglu et~al.(2010)Cambazoglu, Zaragoza, Chapelle, Chen, Liao,
  Zheng, and Degenhardt}]{Cambazoglu_etal_WSDM2010}
B.~Barla Cambazoglu, Hugo Zaragoza, Olivier Chapelle, Jiang Chen, Ciya Liao,
  Zhaohui Zheng, and Jon Degenhardt. 2010.
\newblock Early exit optimizations for additive machine learned ranking
  systems.
\newblock In \emph{Proceedings of the Third ACM International Conference on Web
  Search and Data Mining (WSDM 2010)}, pages 411--420, New York, New York.

\bibitem[{Chen et~al.(2017)Chen, Gallagher, Blanco, and
  Culpepper}]{ChenRuey-Cheng_etal_SIGIR2017a}
Ruey-Cheng Chen, Luke Gallagher, Roi Blanco, and J.~Shane Culpepper. 2017.
\newblock Efficient cost-aware cascade ranking in multi-stage retrieval.
\newblock In \emph{Proceedings of the 40th Annual International ACM SIGIR
  Conference on Research and Development in Information Retrieval (SIGIR
  2017)}, pages 445--454, Tokyo, Japan.

\bibitem[{Cormack et~al.(2009)Cormack, Clarke, and
  {B\"{u}ttcher}}]{cormack2009reciprocal}
Gordon~V. Cormack, Charles L.~A. Clarke, and Stefan {B\"{u}ttcher}. 2009.
\newblock Reciprocal rank fusion outperforms {Condorcet} and individual rank
  learning methods.
\newblock In \emph{Proceedings of the 32nd Annual International ACM SIGIR
  Conference on Research and Development in Information Retrieval (SIGIR
  2009)}, pages 758--759, Boston, Massachusetts.

\bibitem[{Devlin et~al.(2019)Devlin, Chang, Lee, and
  Toutanova}]{devlin2018bert}
Jacob Devlin, Ming-Wei Chang, Kenton Lee, and Kristina Toutanova. 2019.
\newblock {BERT}:\ {P}re-training of deep bidirectional transformers for
  language understanding.
\newblock In \emph{Proceedings of the 2019 Conference of the North {A}merican
  Chapter of the Association for Computational Linguistics:\ Human Language
  Technologies, Volume 1 (Long and Short Papers)}, pages 4171--4186,
  Minneapolis, Minnesota.

\bibitem[{Dietz et~al.(2017)Dietz, Verma, Radlinski, and
  Craswell}]{dietz2017trec}
Laura Dietz, Manisha Verma, Filip Radlinski, and Nick Craswell. 2017.
\newblock {TREC} complex answer retrieval overview.
\newblock In \emph{Proceedings of the Twenty-Sixth Text REtrieval Conference
  (TREC 2017)}.

\bibitem[{Guo et~al.(2016)Guo, Fan, Ai, and Croft}]{guo2016deep}
Jiafeng Guo, Yixing Fan, Qingyao Ai, and W.~Bruce Croft. 2016.
\newblock A deep relevance matching model for ad-hoc retrieval.
\newblock In \emph{Proceedings of the 25th ACM International on Conference on
  Information and Knowledge Management}, pages 55--64, Indianapolis, Indiana.

\bibitem[{Henderson et~al.(2017)Henderson, Al-Rfou, Strope, hsuan Sung, Lukacs,
  Guo, Kumar, Miklos, and Kurzweil}]{Henderson:1705.00652:2017}
Matthew Henderson, Rami Al-Rfou, Brian Strope, Yun hsuan Sung, Laszlo Lukacs,
  Ruiqi Guo, Sanjiv Kumar, Balint Miklos, and Ray Kurzweil. 2017.
\newblock Efficient natural language response suggestion for {Smart} {Reply}.
\newblock \emph{arXiv:1705.00652}.

\bibitem[{Hui et~al.(2018)Hui, Yates, Berberich, and de~Melo}]{hui2018co}
Kai Hui, Andrew Yates, Klaus Berberich, and Gerard de~Melo. 2018.
\newblock {Co-PACRR}: A context-aware neural {IR} model for ad-hoc retrieval.
\newblock In \emph{Proceedings of the Eleventh ACM International Conference on
  Web Search and Data Mining (WSDM 2018)}, pages 279--287, Marina Del Rey,
  California.

\bibitem[{Ji et~al.(2019)Ji, Shao, and Yang}]{Ji:2019:EIN:3308558.3313576}
Shiyu Ji, Jinjin Shao, and Tao Yang. 2019.
\newblock Efficient interaction-based neural ranking with locality sensitive
  hashing.
\newblock In \emph{Proceedings of the 2019 World Wide Web Conference (WWW
  2019)}, pages 2858--2864, San Francisco, California.

\bibitem[{Kashyapi et~al.(2018)Kashyapi, Chatterjee, Ramsdell, and
  Dietz}]{kashyapi2018trema}
Sumanta Kashyapi, Shubham Chatterjee, Jordan Ramsdell, and Laura Dietz. 2018.
\newblock {TREMA-UNH} at {TREC} 2018: Complex answer retrieval and news track.
\newblock In \emph{Proceedings of the Twenty-Seventh Text REtrieval Conference
  (TREC 2018)}.

\bibitem[{Kingma and Ba(2014)}]{kingma2014adam}
Diederik~P. Kingma and Jimmy Ba. 2014.
\newblock Adam: A method for stochastic optimization.
\newblock \emph{arXiv:1412.6980}.

\bibitem[{Li(2011)}]{LiHang_2011}
Hang Li. 2011.
\newblock \emph{Learning to Rank for Information Retrieval and Natural Language
  Processing}.
\newblock Morgan \& Claypool Publishers.

\bibitem[{Lin(2019)}]{lin2019neural}
Jimmy Lin. 2019.
\newblock The neural hype and comparisons against weak baselines.
\newblock In \emph{SIGIR Forum}, volume~52, pages 40--51.

\bibitem[{Lin et~al.(2016)Lin, Crane, Trotman, Callan, Chattopadhyaya, Foley,
  Ingersoll, Macdonald, and Vigna}]{Lin_etal_ECIR2016}
Jimmy Lin, Matt Crane, Andrew Trotman, Jamie Callan, Ishan Chattopadhyaya, John
  Foley, Grant Ingersoll, Craig Macdonald, and Sebastiano Vigna. 2016.
\newblock Toward reproducible baselines: The open-source {IR} reproducibility
  challenge.
\newblock In \emph{Proceedings of the 38th European Conference on Information
  Retrieval (ECIR 2016)}, pages 408--420, Padua, Italy.

\bibitem[{Liu et~al.(2017)Liu, Xiao, Ou, and Si}]{LiuShichen_etal_SIGKDD2017}
Shichen Liu, Fei Xiao, Wenwu Ou, and Luo Si. 2017.
\newblock Cascade ranking for operational e-commerce search.
\newblock In \emph{Proceedings of the 23rd ACM SIGKDD International Conference
  on Knowledge Discovery and Data Mining (SIGKDD 2017)}, pages 1557--1565,
  Halifax, Nova Scotia, Canada.

\bibitem[{Liu(2009)}]{LiuTY_FnTIR2009}
Tie-Yan Liu. 2009.
\newblock Learning to rank for information retrieval.
\newblock \emph{Foundations and Trends in Information Retrieval},
  3(3):225--331.

\bibitem[{MacAvaney et~al.(2019)MacAvaney, Yates, Cohan, and
  Goharian}]{MacAvaney_etal_SIGIR2019}
Sean MacAvaney, Andrew Yates, Arman Cohan, and Nazli Goharian. 2019.
\newblock {CEDR}: Contextualized embeddings for document ranking.
\newblock In \emph{Proceedings of the 42nd Annual International ACM SIGIR
  Conference on Research and Development in Information Retrieval (SIGIR
  2019)}, pages 1101--1104, Paris, France.

\bibitem[{MacAvaney et~al.(2017)MacAvaney, Yates, and
  Hui}]{macavaney2017contextualized}
Sean MacAvaney, Andrew Yates, and Kai Hui. 2017.
\newblock Contextualized {PACRR} for complex answer retrieval.
\newblock In \emph{Proceedings of the Twenty-Sixth Text REtrieval Conference
  (TREC 2017)}.

\bibitem[{Mackenzie et~al.(2018)Mackenzie, Culpepper, Blanco, Crane, Clarke,
  and Lin}]{Mackenzie_etal_WSDM2018}
Joel Mackenzie, Shane Culpepper, Roi Blanco, Matt Crane, Charles Clarke, and
  Jimmy Lin. 2018.
\newblock Query driven algorithm selection in early stage retrieval.
\newblock In \emph{Proceedings of the 11th ACM International Conference on Web
  Search and Data Mining (WSDM 2018)}, pages 396--404, Marina Del Rey,
  California.

\bibitem[{Matveeva et~al.(2006)Matveeva, Burges, Burkard, Laucius, and
  Wong}]{Matveeva_etal_SIGIR2006}
Irina Matveeva, Chris Burges, Timo Burkard, Andy Laucius, and Leon Wong. 2006.
\newblock High accuracy retrieval with multiple nested ranker.
\newblock In \emph{Proceedings of the 29th Annual International ACM SIGIR
  Conference on Research and Development in Information Retrieval (SIGIR
  2006)}, pages 437--444, Seattle, Washington.

\bibitem[{Mitra and Craswell(2019)}]{MitraBhaskar_Craswell_2019}
Bhaskar Mitra and Nick Craswell. 2019.
\newblock An introduction to neural information retrieval.
\newblock \emph{Foundations and Trends in Information Retrieval}, 13(1):1--126.

\bibitem[{Mitra et~al.(2017)Mitra, Diaz, and Craswell}]{mitra2017learning}
Bhaskar Mitra, Fernando Diaz, and Nick Craswell. 2017.
\newblock Learning to match using local and distributed representations of text
  for web search.
\newblock In \emph{Proceedings of the 26th International Conference on World
  Wide Web (WWW 2017)}, pages 1291--1299, Perth, Australia.

\bibitem[{Montague and Aslam(2002)}]{montague2002condorcet}
Mark Montague and Javed~A. Aslam. 2002.
\newblock Condorcet fusion for improved retrieval.
\newblock In \emph{Proceedings of the Eleventh International Conference on
  Information and Knowledge Management (CIKM 2002)}, pages 538--548, McLean,
  Virginia.

\bibitem[{{M\"{u}hleisen} et~al.(2014){M\"{u}hleisen}, Samar, Lin, and
  de~Vries}]{Muhleisen_etal_SIGIR2014}
Hannes {M\"{u}hleisen}, Thaer Samar, Jimmy Lin, and Arjen de~Vries. 2014.
\newblock Old dogs are great at new tricks: Column stores for {IR} prototyping.
\newblock In \emph{Proceedings of the 37th Annual International ACM SIGIR
  Conference on Research and Development in Information Retrieval (SIGIR
  2014)}, pages 863--866, Gold Coast, Australia.

\bibitem[{Nogueira and Cho(2019)}]{nogueira2019passage}
Rodrigo Nogueira and Kyunghyun Cho. 2019.
\newblock Passage re-ranking with {BERT}.
\newblock \emph{arXiv:1901.04085}.

\bibitem[{Onal et~al.(2018)Onal, Zhang, Altingovde, Rahman, Karagoz, Braylan,
  Dang, Chang, Kim, McNamara, Angert, Banner, Khetan, McDonnell, Nguyen, Xu,
  Wallace, de~Rijke, and Lease}]{Onal_etal_IRJ2018}
Kezban~Dilek Onal, Ye~Zhang, Ismail~Sengor Altingovde, Md~Mustafizur Rahman,
  Pinar Karagoz, Alex Braylan, Brandon Dang, Heng-Lu Chang, Henna Kim, Quinten
  McNamara, Aaron Angert, Edward Banner, Vivek Khetan, Tyler McDonnell,
  An~Thanh Nguyen, Dan Xu, Byron~C. Wallace, Maarten de~Rijke, and Matthew
  Lease. 2018.
\newblock Neural information retrieval: At the end of the early years.
\newblock \emph{Information Retrieval}, 21(2--3):111--182.

\bibitem[{Pedersen(2010)}]{Pedersen_SIGIR2010}
Jan Pedersen. 2010.
\newblock Query understanding at {Bing}.
\newblock In \emph{Industry Track Keynote at the 33rd Annual International ACM
  SIGIR Conference on Research and Development in Information Retrieval (SIGIR
  2010)}, Geneva, Switzerland.

\bibitem[{Peters et~al.(2017)Peters, Ammar, Bhagavatula, and
  Power}]{peters2017semi}
Matthew~E. Peters, Waleed Ammar, Chandra Bhagavatula, and Russell Power. 2017.
\newblock Semi-supervised sequence tagging with bidirectional language models.
\newblock \emph{arXiv:1705.00108}.

\bibitem[{Radford et~al.(2018)Radford, Narasimhan, Salimans, and
  Sutskever}]{radford2018improving}
Alec Radford, Karthik Narasimhan, Tim Salimans, and Ilya Sutskever. 2018.
\newblock Improving language understanding by generative pre-training.

\bibitem[{Raffel et~al.(2019)Raffel, Shazeer, Roberts, Lee, Narang, Matena,
  Zhou, Li, and Liu}]{raffel2019exploring}
Colin Raffel, Noam Shazeer, Adam Roberts, Katherine Lee, Sharan Narang, Michael
  Matena, Yanqi Zhou, Wei Li, and Peter~J. Liu. 2019.
\newblock Exploring the limits of transfer learning with a unified text-to-text
  transformer.
\newblock \emph{arXiv:1910.10683}.

\bibitem[{Robertson et~al.(1994)Robertson, Walker, Jones, Hancock-Beaulieu, and
  Gatford}]{robertson1995okapi}
Stephen~E. Robertson, Steve Walker, Susan Jones, Micheline Hancock-Beaulieu,
  and Mike Gatford. 1994.
\newblock Okapi at {TREC-3}.
\newblock In \emph{Proceedings of the 3rd Text REtrieval Conference (TREC-3)},
  pages 109--126, Gaithersburg, Maryland.

\bibitem[{Viola and Jones(2004)}]{Viola_Jones_208}
Paul Viola and Michael~J. Jones. 2004.
\newblock Robust real-time face detection.
\newblock \emph{International Journal of Computer Vision}, 57:137--154.

\bibitem[{Voorhees(2002)}]{Voorhees_CLEF2002}
Ellen~M. Voorhees. 2002.
\newblock The philosophy of information retrieval evaluation.
\newblock In \emph{Evaluation of Cross-Language Information Retrieval Systems:
  Second Workshop of the Cross-Language Evaluation Forum, Lecture Notes in
  Computer Science Volume 2406}, pages 355--370.

\bibitem[{Wang et~al.(2011)Wang, Lin, and Metzler}]{Wang_etal_SIGIR2011}
Lidan Wang, Jimmy Lin, and Donald Metzler. 2011.
\newblock A cascade ranking model for efficient ranked retrieval.
\newblock In \emph{Proceedings of the 34th Annual International ACM SIGIR
  Conference on Research and Development in Information Retrieval (SIGIR
  2011)}, pages 105--114, Beijing, China.

\bibitem[{Xiong et~al.(2017)Xiong, Dai, Callan, Liu, and Power}]{xiong2017end}
Chenyan Xiong, Zhuyun Dai, Jamie Callan, Zhiyuan Liu, and Russell Power. 2017.
\newblock End-to-end neural ad-hoc ranking with kernel pooling.
\newblock In \emph{Proceedings of the 40th International ACM SIGIR Conference
  on Research and Development in Information Retrieval (SIGIR 2017)}, pages
  55--64, Tokyo, Japan.

\bibitem[{Xu et~al.(2012)Xu, Weinberger, and
  Chapelle}]{XuZhixiang_etal_ICML2012}
Zhixiang~Eddie Xu, Kilian~Q. Weinberger, and Olivier Chapelle. 2012.
\newblock The greedy miser: Learning under test-time budgets.
\newblock In \emph{Proceedings of the 29th International Conference on Machine
  Learning (ICML 2012)}, Edinburgh, Scotland.

\bibitem[{Yang et~al.(2017)Yang, Fang, and Lin}]{Yang_etal_SIGIR2017}
Peilin Yang, Hui Fang, and Jimmy Lin. 2017.
\newblock {Anserini}: Enabling the use of {Lucene} for information retrieval
  research.
\newblock In \emph{Proceedings of the 40th Annual International ACM SIGIR
  Conference on Research and Development in Information Retrieval (SIGIR
  2017)}, pages 1253--1256, Tokyo, Japan.

\bibitem[{Yang et~al.(2018)Yang, Fang, and Lin}]{Yang_etal_JDIQ2018}
Peilin Yang, Hui Fang, and Jimmy Lin. 2018.
\newblock {Anserini}: Reproducible ranking baselines using {Lucene}.
\newblock \emph{Journal of Data and Information Quality}, 10(4):Article 16.

\bibitem[{Yang et~al.(2019{\natexlab{a}})Yang, Lu, Yang, and
  Lin}]{Yang_etal_arXiv2019hype}
Wei Yang, Kuang Lu, Peilin Yang, and Jimmy Lin. 2019{\natexlab{a}}.
\newblock Critically examining the ``neural hype'':\ weak baselines and the
  additivity of effectiveness gains from neural ranking models.
\newblock In \emph{Proceedings of the 42nd Annual International ACM SIGIR
  Conference on Research and Development in Information Retrieval (SIGIR
  2019)}, pages 1129--1132, Paris, France.

\bibitem[{Yang et~al.(2019{\natexlab{b}})Yang, Xie, Lin, Li, Tan, Xiong, Li,
  and Lin}]{Yang_etal_NAACL2019demo}
Wei Yang, Yuqing Xie, Aileen Lin, Xingyu Li, Luchen Tan, Kun Xiong, Ming Li,
  and Jimmy Lin. 2019{\natexlab{b}}.
\newblock End-to-end open-domain question answering with {BERTserini}.
\newblock In \emph{Proceedings of the 2019 Conference of the North American
  Chapter of the Association for Computational Linguistics (Demonstrations)},
  pages 72--77, Minneapolis, Minnesota.

\bibitem[{Yang et~al.(2019{\natexlab{c}})Yang, Zhang, and
  Lin}]{Yang_etal_arXiv2019b}
Wei Yang, Haotian Zhang, and Jimmy Lin. 2019{\natexlab{c}}.
\newblock Simple applications of {BERT} for ad hoc document retrieval.
\newblock \emph{arXiv:1903.10972}.

\bibitem[{Yilmaz et~al.(2019)Yilmaz, Yang, Zhang, and
  Lin}]{Yilmaz_etal_EMNLP2019}
Zeynep~Akkalyoncu Yilmaz, Wei Yang, Haotian Zhang, and Jimmy Lin. 2019.
\newblock Cross-domain modeling of sentence-level evidence for document
  retrieval.
\newblock In \emph{Proceedings of the 2019 Conference on Empirical Methods in
  Natural Language Processing}.

\bibitem[{Zamani et~al.(2018)Zamani, Dehghani, Croft, Learned-Miller, and
  Kamps}]{Zamani:2018:NRN:3269206.3271800}
Hamed Zamani, Mostafa Dehghani, W.~Bruce Croft, Erik Learned-Miller, and Jaap
  Kamps. 2018.
\newblock From neural re-ranking to neural ranking: Learning a sparse
  representation for inverted indexing.
\newblock In \emph{Proceedings of the 27th ACM International Conference on
  Information and Knowledge Management (CIKM 2018)}, pages 497--506, Torino,
  Italy.

\end{thebibliography}
\bibliographystyle{acl_natbib}

\end{document}